# Roles of scattered and ballistic photons in imaging through scattering media: a deep learning-based study

Xuyu Zhang[1,2], Jingjing Gao[2,3], Yu Gan[2,3], Chunyuan Song[2,3], Dawei Zhang[1,*], Songlin Zhuang[1], Shensheng Han[2,3,4], Puxiang Lai[5,6,7*], and Honglin Liu[2,3,6,*]

[1]*School of Optical-Electrical and Computer Engineering, University of Shanghai for Science and Technology, Shanghai 200093, China*
[2]*Key Laboratory for Quantum Optics, Shanghai Institute of Optics and Fine Mechanics, Chinese Academy of Sciences, Shanghai 201800, China*
[3]*Center of Materials Science and Optoelectronics Engineering, University of Chinese Academy of Science, Beijing 100049, China*
[4]*Hangzhou Institute for Advanced study, University of Chinese Academy of Sciences, Hangzhou 310024, China*
[5]*Department of Biomedical Engineering, The Hong Kong Polytechnic University, Hong Kong SAR, China*
[6]*Hong Kong Polytechnic University Shenzhen Research Institute, Shenzhen 518000, China*
[7]*Photonics Research Institute, The Hong Kong Polytechnic University, Hong Kong SAR, China*

*[dwzhang@usst.edu.cn](mailto:dwzhang@usst.edu.cn), [puxiang.lai@polyu.edu.hk](mailto:puxiang.lai@polyu.edu.hk), and [hlliu4@hotmail.com](mailto:hlliu4@hotmail.com)*

**Abstract**

Scattering of light in complex media scrambles optical wavefronts and breaks the principles of conventional imaging methods. For decades, researchers have endeavored to conquer the problem by inventing approaches such as adaptive optics, iterative wavefront shaping, and transmission matrix measurement. That said, imaging through/into thick scattering media remains challenging to date. With the rapid development of computing power, deep learning has been introduced and shown potentials to reconstruct target information through complex media or from rough surfaces. But it also fails once coming to optically thick media where ballistic photons become negligible. Here, instead of treating deep learning only as an image extraction method, whose best-selling advantage is to avoid complicate physical models, we exploit it as a tool to explore the underlying physical principles. By adjusting the weights of ballistic and scattered photons through a random phasemask, it is found that although deep learning can extract images from both scattered and ballistic light, the mechanisms are different: scattering may function as an encryption key and decryption from scattered light is key sensitive, while extraction from ballistic light is stable. Based on this finding, it is hypothesized and experimentally confirmed that the foundation of the generalization capability of trained neural networks for different diffusers can trace back to the contribution of ballistic photons, even though their weights of photon counting in detection are not that significant. Moreover, the study may pave an avenue for using deep learning as a probe in exploring the unknown physical principles in various fields.

## 1. Introduction

Imaging through scattering media has always been an important topic in the community of optical imaging. Traditionally, to achieve high-resolution imaging, three main strategies are usually adopted. One is to select or extract ballistic photons such as in time-of-flight methods [3,4], optical coherence tomography (OCT) [5-8], as well as confocal/multi-photon microscopies [9-12]. The second is to compensate the phase scrambling of scattered photons such as in optical phase conjugation [13-17], wavefront shaping [18-20], and scattering matrix measurement [21-23]. Speckle deconvolution and speckle autocorrelation imaging [24-26] belong to the third category, in which intrinsic statistical properties of scattered photons and optical memory effect [27] are utilized, so that the field of view (FOV) is inversely proportional to the thickness of the disorder medium. For all these high-resolution approaches, the penetration thickness is mostly limited within the optical diffusion limit [28] to date, although some studies were believed to be able to break the thickness bottleneck [Refs].

In the past few years, rapid development of artificial intelligence and computing power has enabled wide applications of deep learning in different fields [29-35]. Compared with the aforementioned methods, deep learning-based approach is free from complex experimental alignment, formula derivation, or reconstruction algorithm, but relies on an end-to-end mapping. The mapping, usually represented by neural networks, is trained by thousands of or more pairs of inputs and outputs. Note that, however, no specific mathematic equations or physical models are aimed at during this process and the neural network functions like a black box with only inputs and outputs exposed to users, which is its key advantage but also major shortcoming. It's difficult to justify to what extent a neural network can mimic an unknown scattering system, and how reliable the predictions are. In traditional fields, like image recognition [29-31], language processing [32], and strategy selection [33, 34], the internal principle is known. The main strengths of deep learning are to enhance speed and efficiency, seek optimal solutions for ill-posed problems, and so on. However, for imaging through scattering media, without solid foundation and boundary in physics, it is hardly possible to apply deep learning in practical applications with sufficient confidence.

In 2016, Horisaki *et al.* introduced machine learning based on nonlinear support vector regression to predict face targets hidden behind acrylic resin plates [36], although the reconstructed faces were not accurate and lacked details. Moreover, the trained network still reconstructed a face from a speckle pattern when a non-facial target appeared in a testing set. In 2017, Lyu *et al.* proposed a deep learning method to retrieve images of written digits and letters through a 3-mm thick white polystyrene slab that had an optical thickness of 13.4 mean free paths [37]. In 2018, S. Li *et al.* proposed a convolutional neural network, named as diffNet, to image through glass diffusers. It worked well and showed some generalization capability in predicting objects of unseen types through a seen 600-grit diffuser (DG10-600-MD, Thorlabs) [38], but the network failed for a diffuser that was more diffusive (220-grit, 45-653, Edmund). In the same year, Y. Li *et al.* developed a convolutional neural network (CNN) with improved generalization capability in predicting objects through unseen diffusers [39]. The network was trained by speckle patterns captured through a set of 220-grit diffusers (DG10-220, Thorlabs), which shared the same macroscopic parameters.

The topic has attracted intensive interests and witnessed encouraging development in recent years [37-39], and it has also been demonstrated that simulation data can be used for network training [39-41]. That said, it remains challenging to see or image through thick scattering media; empirically, it seems that deep learning functions only when there are significant amounts of ballistic or quasi-ballistic photons. It is hence interesting to explore whether the deep learning-based approaches can retrieve effective information from scattered photons. If yes, why increased thickness of the scattering medium poses challenges? If not, what are the roles of scattered and ballistic photons in this process?

Aiming to answer these questions, a neural network is used as a probe rather than a simple image extraction method to explore the underlying physical principles. A set of experiments and simulations are performed in this study. In experiment, instead of using different diffusers, we obtain the training data from different regions of a homemade 220-grit diffuser, achieving successful imaging through untrained, *i.e.*, unseen, or partially trained areas of the diffuser. In simulation, the same experimental configuration is employed. Benefitting from the adjustable phase distribution while preserving the macroscopic parameters of simulated diffusers, we could control the weights of ballistic and scattered photons. In simulation, influences of factors like sample thickness, memory effect range determined FOV, and mechanical vibration are shielded, so that the study is focused on the ratio between ballistic and scattered photons. The simulation results suggest that deep learning can extract information from both scattered and ballistic photons under coherent illumination, but via different mechanisms: scattering may function as an encryption key and decryption from scattered light is key sensitive, while extraction from ballistic light is stable. The hypothesis is experimentally confirmed that the foundation of the generalization capability of trained network for unseen scattering media may trace back to the contribution of ballistic photons, even though their weights of photon counting in detection are not that significant. These findings set a boundary of scattering for deep learning to function through turbid media and provide guidelines to optimize the experimental framework for imaging prediction.

## 2. Methods

### 2.1 Experimental Setup

As shown in Fig. 1, a 532 nm solid-state laser (MGL-III-532-200mW, Changchun New Industries Optoelectronics Tech) is expanded and collimated onto a digital micromirror device (DMD, V-7001 VIS, ViALUX) that has a pixel pitch of 13.7 μm. The reflected beam illuminates a homemade 220-grit ground glass diffuser. An iris with a diameter of 5 mm is placed right after the diffuser to create a tunable optical window. The transmitted light is collected by a digital camera (DCU224M, Thorlabs), whose resolution is 1280×1024 with a pixel pitch of 4.65 μm. The distances from the beam expander to the DMD, from DMD to the diffuser, and from the diffuser to the camera are $z_1 = 15\ cm$, $z_2 = 16\ cm$, and $z_3 = 10\ cm$, respectively. Handwritten digits from the MNIST database [42] are used as the objects, which are reshaped into 64×64 arrays and loaded on the central 64×64 pixels of the DMD in subsequence. The diameter of collimated beam on the DMD is 5 mm, sufficiently covering the whole region of the digits.

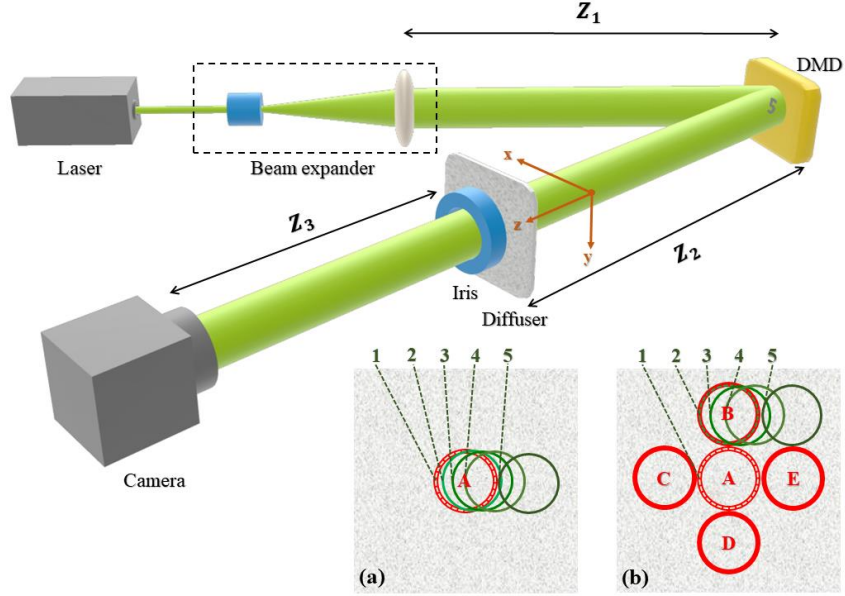

Fig. 1. Schematic of the experimental setup. Next to the diffuser is the coordinate system. DMD: digital micromirror device. Inserts (a) and (b) denote the position setting to acquire the training and testing data in Tests I and II, respectively. In (a), the training data were obtained from Position A, and the testing data were from Positions 1-5. In (b), the training data were obtained from Positions A-E, and the testing data were from Positions 1-5. Note that Position 1 overlapped with Position A in both scenes.

Two different settings were adopted to arrange the positions to acquire the neural network training and testing data, as indicated in inserts (a) and (b) in Fig. 1. The diameters of the circles were the same with that of the iris. In Test I (Fig. 1a), 20,000 pairs of training data were obtained at Position A, as represented by the red circle. Data from 5 different positions, denoted by the green circles 1 to 5, were used for testing. Note that Position 1 overlapped with Position A, and Positions 2, 3, 4, and 5 were shifted 10 μm, 40 μm, 100 μm, and 5000 μm, respectively, along the x axis direction from Position 1. In Test II (Fig. 1b), 20,000 pairs of data obtained from Positions A to E, with 4,000 pairs at each position, were used to train the neural network, while data from Positions 1-5 were for test. Note that Positions 1 and A overlapped. When the diffuser was translated by 5000 μm along the y axis, Positions 2 and B overlapped. Positions 3, 4, and 5 were shifted from Position 2 along the x axis by 40 μm, 100 μm, and 5000 μm, respectively.

According to the Van Cittert-Zernike theorem [43], the spatial coherence length, *i.e.*, the averaged speckle size on the camera plane, $l_c = \frac{\lambda z_3}{D}$, where $\lambda$ is the optical wavelength, $D$ is the diameter of the iris. By substituting the parameters used in our experiment, the averaged speckle size was 10.64 μm, about 2-3 times of the pitch of camera pixels. Therefore, the speckle distribution in our experiment could be recorded with sufficient spatial sampling rate.

To increase the speed of training, the central areas of recorded CCD intensity patterns 512×512 were cropped into 256×256 arrays. To quantify the isoplanatic range of the diffuser (phasemask), we used a point source, generated by 3×3 binning pixels on the DMD, to illuminate the diffuser (phasemask) and record the speckle pattern at each displacement. The correlation coefficient of speckle patterns with respect to the one at $\Delta x = 0$ are calculated. The full width at half maximum of the fitted curve of the coefficients denotes the averaged grain size of the diffuser.

## 2.2 Theoretical models

If the optical field incident on the scattering layer is $E_0$, at the output the layer, the field turns to be

$$E = \alpha E_0 + \beta T E_0, \tag{1}$$

where $\alpha$ and $\beta$ are the coefficients denoting the weights of ballistic and scattered photons, respectively, and $T$ is the scattering transmission matrix of the layer. Note that for simplicity, the propagation factor has been neglected in the equation. If the scattering layer is thin, and there is residual ballistic field, then $\alpha > 0$. The speckle pattern on the image plane is

$$I = |\alpha E_0 + \beta T E_0|^2 = \alpha^2|E_0|^2 + \alpha\beta E_0 E_0^* T^* + \alpha\beta E_0^* T E_0 + \beta^2|T E_0|^2, \tag{2}$$

where $|E_0|^2$ is a diffraction pattern of the object, $|TE_0|^2$ denotes a speckle pattern of the scrambled wavefront, and the two cross terms $E_0 E_0^* T^*$ and $E_0^* T E_0$ also have speckle appearances. Usually, $\alpha$ is small for scattering media, thus the diffraction pattern is submerged in speckles. However, the diffraction pattern is constant even when there are motions or vibrations of the scattering layer, as it is independent from the scattering TM.

With no surprise, object information can be easily extracted from the diffraction patterns with deep learning or other methods. But when there are no ballistic photons and only $|TE_0|^2$ term remains, if the trained deep learning network could still extract information from the intensity patterns, intuitively, the network should be sensitive to the motions of the layer. It would fail to predict through an untrained/unseen region. When both ballistic and scattered photons exist, the object information can be extracted by the trained network. It is hence natural to hypothesize that the more is the portion of the ballistic photons, the more robust is the network to the motions of the layer. Moreover, the network could be trained to accommodate the fluctuations of the ballistic portion. To demonstrate these hypotheses, we designed several simulations to train and test a CNN. And to quantify the quality of reconstructed images, Pearson correlation coefficient (PCC) was adopted to measure the similarity between the reconstructed images and the corresponding objects, that is

$$\rho_{X,Y} = \frac{cov(X,Y)}{\sigma_X \sigma_Y}, \tag{3}$$

where $cov(X,Y)$ is the covariance of two variables $X$ and $Y$, $\sigma_X$ and $\sigma_Y$ are their standard deviations.

### 2.3 CNN Implementation

The CNN used in this study is a UNet network, which has been demonstrated to be suitable for image segmentation, restoration, enhancement [44-46]. The structure of our modified UNet network is shown in Fig. 2. On the left is the contraction path, which uses down-sampling and convolution modules to extract the features of different scales. The expansion path on the right uses up-sampling and convolution modules to restore the scale and integrate the previous features to gradually restore the image. The gray arrows indicate jump connections. Their function is to merge the features on the contraction path into the features on the expansion path at the same scale, which can provide better constraints on the expansion path and make it easier for the network to output the desired results. The input of the CNN is preprocessed 256×256 speckle patterns. Then we define a repeated convolution kernel of size 3×3, and each convolution kernel forms a convolution layer with batch normalization (BN) and a rectified liner unit (ReLU). Next, each convolution layer and a maxpooling operation form a down-sampling operation. The whole contraction path consists of four down-sampling operations. Then it transits from a feature layer to an extension path. The expansion path consists of four up-sampling operations, and each up-sampling operation consists of a deconvolution layer and a feature merging operation. The last is the output layer where Sigmoid is used as the activation function. The outputs of trained UNet network are the reconstructed images with a size of 256×256.

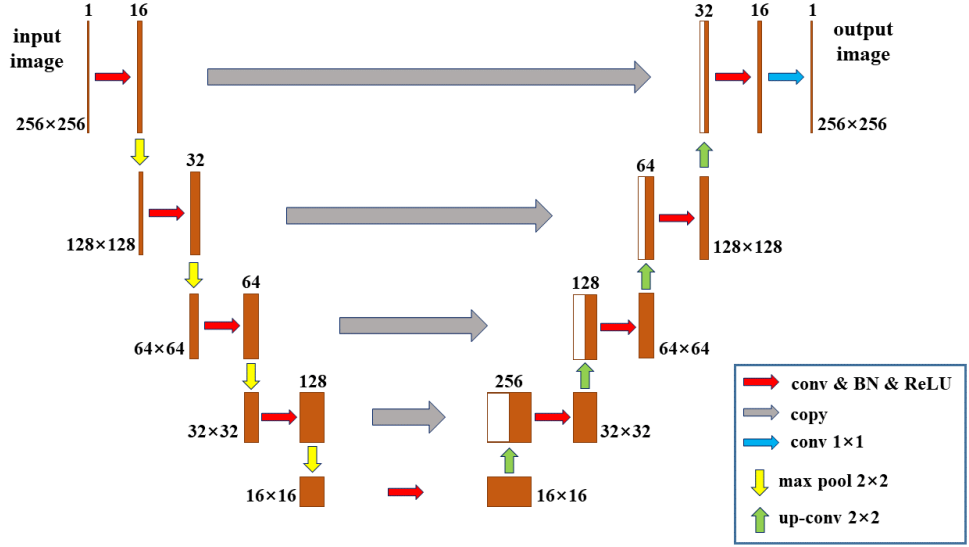

Fig. 2. Schematic of the UNet network. The horizontal objects have the same size, and the vertical ones have the same bit depth.

Here, mean-square error (MSE) is used as the loss function in the training process.

$$MSE = \frac{1}{n}\sum_i^n (y_i - \overline{y_i})^2, \qquad (4)$$

where $n$ represents the total number of speckle patterns and label objects used for training, $y_i$ represents the predicted value of an object, and $\overline{y_i}$ represents the label value of the object.

The UNet network uses Python compiler based on Keras/Tensorflow 2.0 Library, and the GPU edition of the training UNet is NVIDIA RTX 3060 laptop. The total number of training epochs is 50, and the learning rate at the beginning is set as $2\times10^{-4}$. After 5 epochs, if the loss value doesn't decrease, the learning rate will be adjusted to one tenth of the previous one until the learning rate is reduced to $2\times10^{-6}$. When the loss value does not decrease after 10 epochs, the training will be terminated. The averaged training duration of each epoch is 130 s.

### 2.4 Simulations

In simulation based on wave optics, we adopted the same configuration and parameters as the experiments. A diffuser with an array of 3,000×3,000 was generated in advance, with a pitch size of 5 μm. Only a segment of the array, which was 1,000×1,000 that corresponded to the 5 mm diameter iris, was selected as the effective zone to calculate a speckle pattern.

The statistical properties of the diffuser can also be quantified by macroscopic parameters, *i.e.*, the spatial coherence length and the standard derivation of height [47]. Their typical values for a 220-grit diffuser are 36 μm and 1.6 μm, respectively. Supplement I is the code to generate a simulated diffuser, *i.e.*, a random phasemask. By changing coefficient C in the code, we can control the portion of ballistic photons $\eta$ after transmitting through the phasemask, while preserving the spatial coherence length of the phasemask at small $\eta$: when $\eta = 0$, all photons are scattered; when $\eta = 1$, all photons are ballistic and there are no scattering events occurring on the phasemask, which degenerates into a flat phase plate. In this study, the network was trained and tested with simulated speckle patterns of $\eta = 0, 0.1, 0.3, 0.5, 0.7, 0.9$, respectively.

**Simulation I**: $\eta = 0$ was adopted. Similar to the settings in experiment, in the first group, UNet was first trained by 20,000 input-output pairs captured at Position A of the phasemask, then was tested by speckle patterns obtained from Positions 1 to 9, which were horizontally shifted by 0 μm, 10 μm, 35 μm, 40 μm, 100 μm, 1000 μm, 2500 μm, 3500 μm, and 5000 μm, respectively, from Position A. In the second group, the network was trained by 20,000 input-output pairs captured at Positions A to E with 4,000 pairs at each position and then tested by speckle patterns from Positions 1 to 9.

**Simulation II**: $\eta = 0.1, 0.3, 0.5, 0.7, 0.9$ were adopted. At each value, 20,000 speckle patterns from only one position were collected to train the UNet. Data from 6 regions with horizontal shifts of 0 μm, 10 μm, 35 μm, 40 μm, 100 μm, and 5000 μm from that position were respectively used for network testing.

### 3. Simulation and experimental results

Fig. 3 shows the experimental results for the two training and testing settings. In Test I, the quality of reconstructed image was best at Position 1, as the data for network training and testing were from the same position. When the testing data were from Position 2, which was moved by 10 μm from Position A/1, the reconstruction quality deteriorated, but still acceptable. When the positions of testing data acquisition were moved further to Positions 3 and 4, the reconstruction quality became very poor as these two regions were visited or "seen" by the training test for only a small portion. When it was moved to Position 5 that had no overlapping with Position A at all, the reconstructed image can no longer be correlated with the ground truth. In Test II, images can be well reconstructed for both Positions 1 and 2. For positions that were further away from Positions B, *i.e.*, Positions 3-5, the object can still be visually recognized, although the quality was reduced and Position 5 had no overlapping with Positions A-E that were covered in the network training. The difference between these two groups of experiment indicates the network in Test II is less sensitive to position shifts, which agrees with Ref. [36] where different diffusers of the same type rather than different areas of the same diffuser were used to acquire training data. The trend could be seen more clearly if the PCC with the corresponding ground truth is plotted as a function of displacement along the x direction (Fig. 3b). For Test I, the PCC reduces fast with increased displacement and drops to below 0.3 when the displacement is 5000 μm (corresponding to Position 5). For Test II, the PCC reduces when it is moved off the center but remains relatively constant no matter the testing region is covered by the training data or not. The spatial coherence length of the diffuser, *i.e.*, the averaged grain size, is 34 μm (Fig. 3c), which is consistent with the fast decay of PCC around 40 μm displacement. Note that from Fig. 3c, it is suggested that there is significant portion of residual ballistic light since the tail of the cross-correlation coefficient curve is around 0.4. Nevertheless, with training data acquired from 5 positions, the generalization capability of the UNet network has been considerably improved. But whether the origin of such improvement comes from better adaptability to scattered light is still unclear.

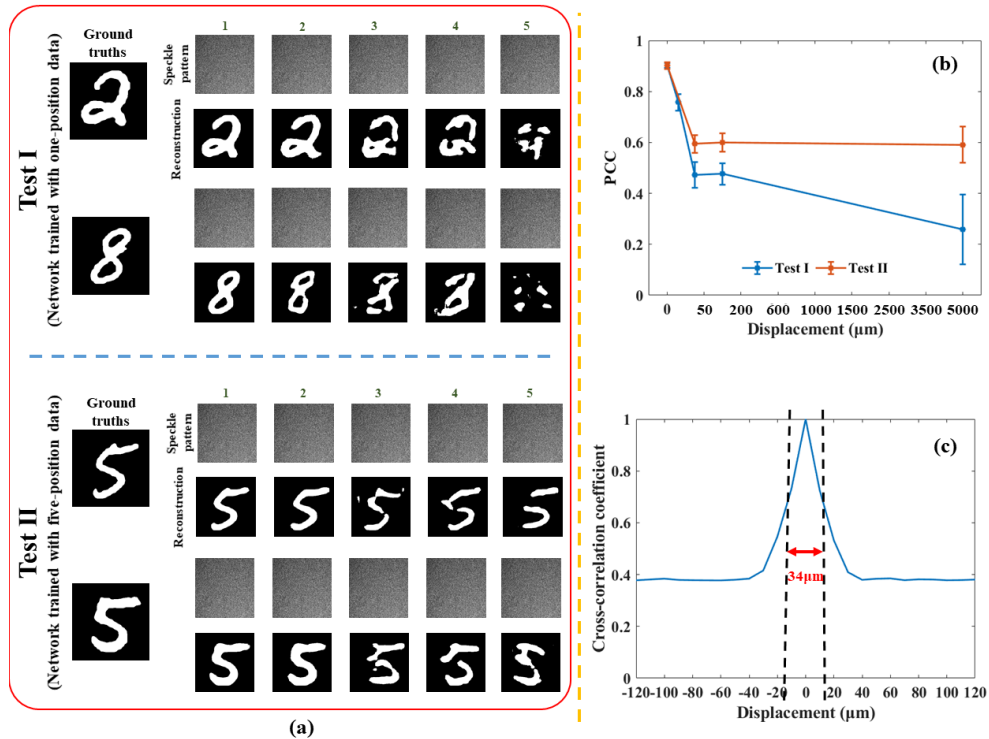

Fig. 3. Experimental results. (a) Reconstructed images through a homemade diffuser in Tests I and II. (b) Averaged PCC of 10 reconstructed images at different positions in both Tests. At large displacements, the reconstructed images of Tests II have better quality than those of Test I. Due to the non-uniform distributed displacements to reflect the whole trend, a non-uniform abscissa axis is adopted, instead of a linear or logarithm one. (c) The cross-correlation coefficient curve of the diffuser at different displacements along the x direction. The estimated speckle size is ~34 μm, as measured by full-width at half maximum.

Aiming to find out the answer, in Simulation I, the influence of ballistic photons is shielded ($\eta = 0$), but only scattered photons are considered in the Unet network. The results are shown in Fig. 4. The spatial coherent length of the random phasemask is 36 μm, agreeing with the selected parameters in the code but is slightly different from the homemade diffuser. As seen, images can be well restored at Positions 1 and 2, but the image quality drops abruptly around 36 μm displacements from the training region. In both groups, the trained networks cannot reconstruct the objects once the displacement is beyond the spatial coherent length. This suggests that even the network is trained by data acquired from 5 different positions (the second group), the generalization capability of the network sees no obvious enhancement if only scattered light is involved. In other words, the network generalization capability observed in experiments (Fig. 3) might be due to the contribution of ballistic light.

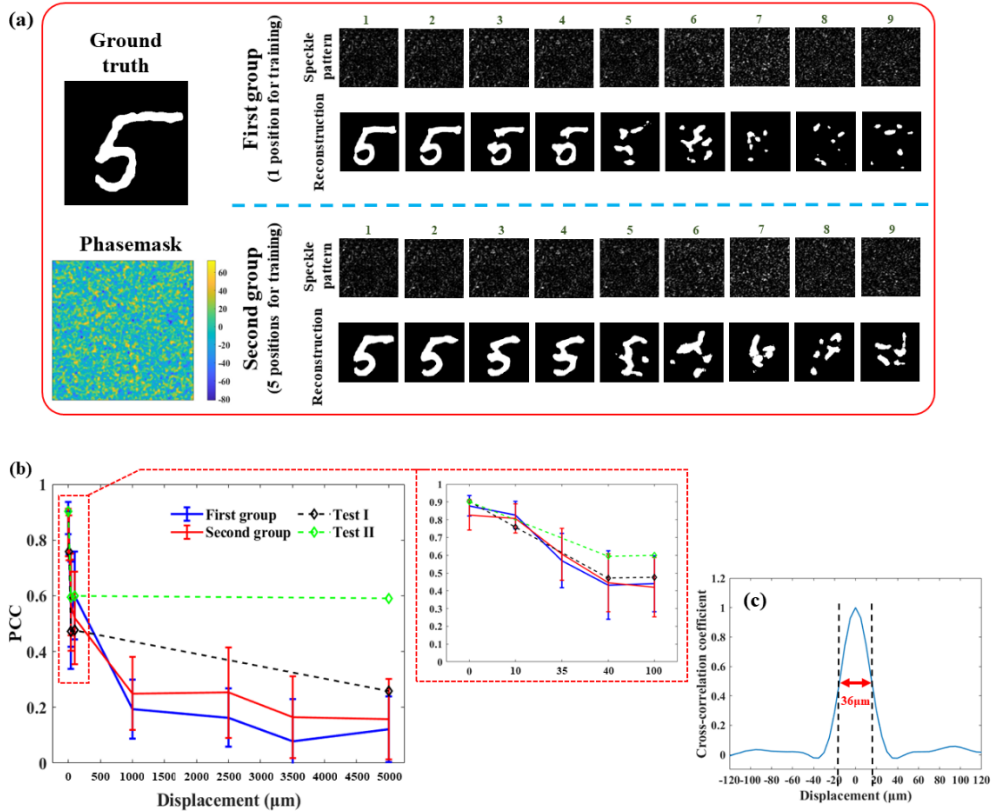

Fig. 4. Results of Simulation I where no ballistic light is involved. (a) The object to be displayed on the SLM (ground truth; top left), the applied phasemask (bottom left), the generated speckle patterns as well as the corresponding reconstructed images at 9 different testing regions that are respectively 0 μm, 10 μm, 35 μm, 40 μm, 100 μm, 1000 μm, 2500 μm, 3500 μm and 5000 μm shifted from the central position where data for network training are acquired. In the first group (top right), the training data are from 1 position (Position A in Fig. 1a); in the second group (bottom right), the training data are from 5 positions (Positions A-E in Fig. 1b). (b) Averaged PCC of the reconstructed images as a function of displacement for two groups of simulation. Experimental results are also included for comparison. The insert shows the zoom-in of the dash rectangle, which includes the displacements of 0 μm, 10 μm, 35 μm, 40 μm, and 100 μm. Again, the abscissa axis is non-uniform. (c) The cross-correlation coefficient curve of the phasemask as a function of displacements along the x axis. The tail level is around 0, confirming the absence of ballistic light in Simulation I. The corresponding speckle grain size is 36 μm.

To further confirm the hypothesis, Simulation II investigates the influence of ballistic light on the generalization capability of trained networks. The corresponding results (Fig. 5) show that as the weight of ballistic light increases, the network is more and more insensitive to displacements along the x direction. When $\eta = 0.9$, the corresponding light pattern (Fig. 5aV) becomes rather smooth, and speckle grains act like just noise. In this situation, the network trained with data acquired from one position can predict the object reliably with testing data from completely unseen regions (Fig. 5bV), demonstrating excellent generalization capability.

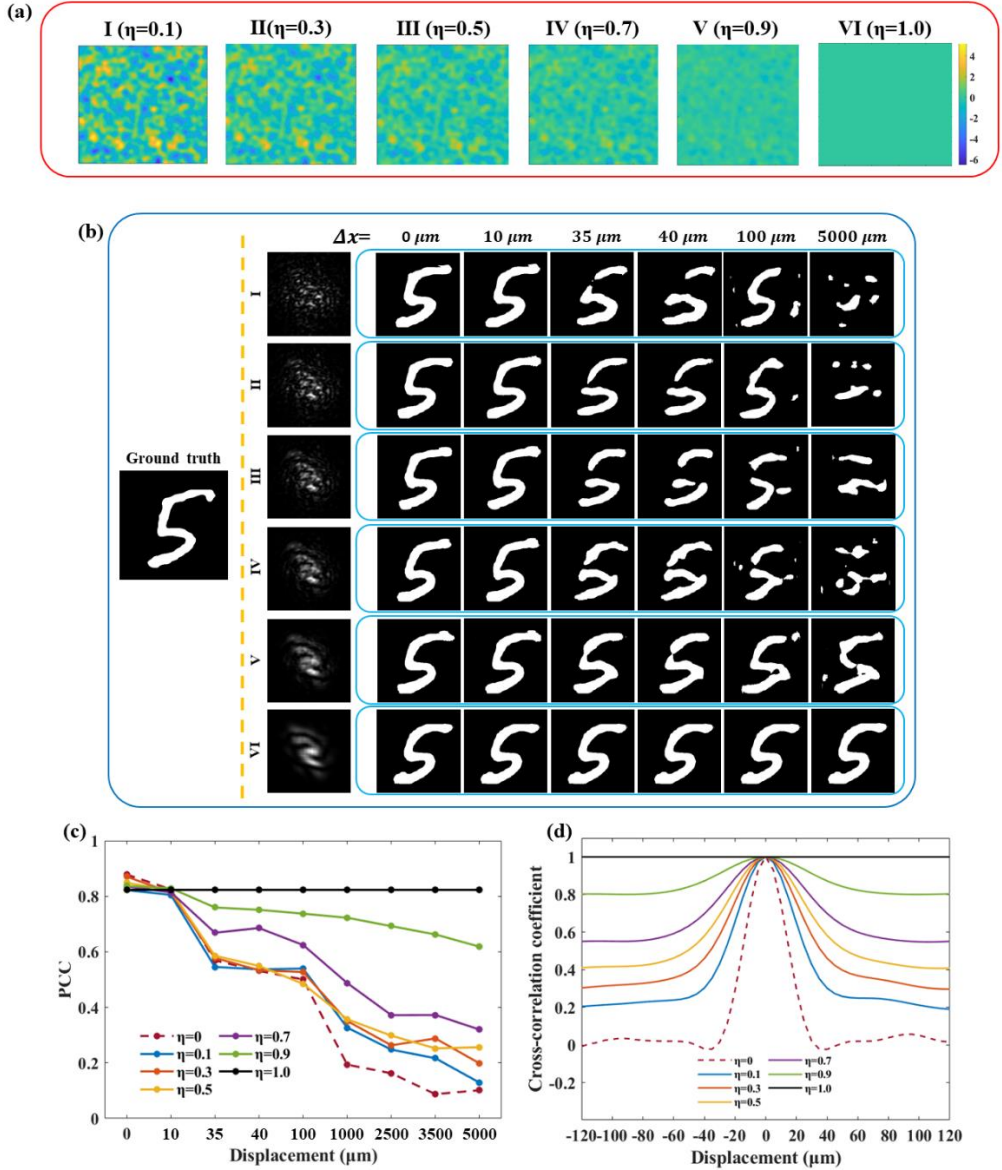

Fig. 5. Results of Simulation II when different weights of ballistic light are involved. (a) The applied phasemasks corresponding to η=0.1, 0.3, 0.5, 0.7, 0.9 and 1, respectively. (b) The object to be displayed on the SLM (ground truth; on the left of the orange dashed line), and the reconstructed images with different weights of ballistic light at different displacements. Rows I, II, III, IV, V, and VI correspond to η=0.1, 0.3, 0.5, 0.7, 0.9, and 1, respectively, and the first column shows the recorded speckle patterns of the digit "5" at $\Delta x = 0\ \mu m$. (c, d) The PCC and cross-correlation coefficient for phasemasks as a function of displacement for different weights of ballistic light.

The trend is more straightforward from the PCC curves (Fig. 5c), where the performance becomes less sensitive to displacement, hence increased generalization capability, with larger portions of ballistic photons. At $\Delta x = 0\ \mu m$, image qualities of different weights of ballistic photons are almost the same within the error bound (comparing to $\eta = 1$, the slightly biased better image qualities for $\eta = 0, 0.1, 0.3, 0.5, 0.7, 0.9$ are caused by propagation redirection of high frequency components on the phasemask and thus been detected by the camera), which is an explicit demonstration that information doesn't lose after scattering. It also demonstrates that a trained network can fully extract the information no matter what pattern it is encoded in. In other words, the deteriorated image qualities at non-zero displacements are caused only by the mismatches between the encryption and the decryption keys, but not the information extraction efficiencies from different components. Currently, all proposed neural networks focus on extracting target features of different scales on their different layers. They don't differentiate scattered or ballistic photons. Since the final extracted information is the same, there shouldn't be significant differences in extracted features or patterns. That suggests that there is little chance to differentiate scattered and ballistic photons inside the neural networks.

From the cross-correlation coefficient curves (Fig. 5d), the spatial coherent length of the phasemask is stable when $\eta$ is small, but grows faster and faster as $\eta$ increases. Also, the level of the curve tails is closely related with the weight of ballistic light, generally proportional to $\eta^2$. But there are discrepancies between the tails with $\eta^2$, especially for smaller $\eta$. The phenomenon is associated with the way $\eta$ is defined: it measures the weight of ballistic photons on the output surface of phasemask but not those entering the detector. Note that diffused photons travel along random directions and hence have lower possibilities to be sensed by the detector with a limited aperture. Therefore, the actual weight of ballistic photons among all photons that enter the detector is higher than $\eta^2$, and the smaller is $\eta$, the larger is the discrepancy. For example, when $\eta = 0.1$, the tail level in Fig. 5d is 0.2, being considerably higher than $0.1^2=0.01$; but when $\eta = 0.9$, the tail level becomes ~0.8, which is very close to $0.9^2=0.81$. Nevertheless, from the tail level, we can roughly estimate the weight of ballistic photons on the detection plane. For instance, in our experiment the tail level is ~0.37 (Fig. 3c), suggesting a $\eta \approx 0.61$. Although this value of estimation is not identical to that reported in Ref. [36] ($\eta \approx 0.55$), both validate the detection of significant portions of ballistic light through the ground glass diffuser used in experiment.

The above results suggest that the learning-based performance of imaging reconstruction through a scattering medium, for example as measured by PCC, can be improved by filtering out some scattered photons and/or increasing the weight of ballistic photons. A simple way to increase the weight of ballistic photons is to increase $z_3$ in Fig. 1, the distance from the scattering medium to the detector, considering a limited aperture for the detector and diffused photon propagate in all directions. Experimentally, we increased $z_3$ from 5 to 10 and 20 cm, selected a 256×256 (for $z_3 = 5\ cm$) or 1024×1024 (for the other two cases) array on the camera to ensure the same number of independent speckles on the detection plane, and repeated Tests I and II following the settings in Fig. 1. By doing so, according to what was reasoned above, the weight of ballistic photons was enhanced, which can be confirmed in Fig. 6a: the level of the curve tails was increased from ~0.31 to ~0.37 and ~0.41, which corresponds to ratio of ballistic photons from ~0.56 to ~0.61 and ~0.64. The performance of image reconstruction, as measured by the PCC with respect to the ground truth, is also improved as forecasted (Fig. 6b).

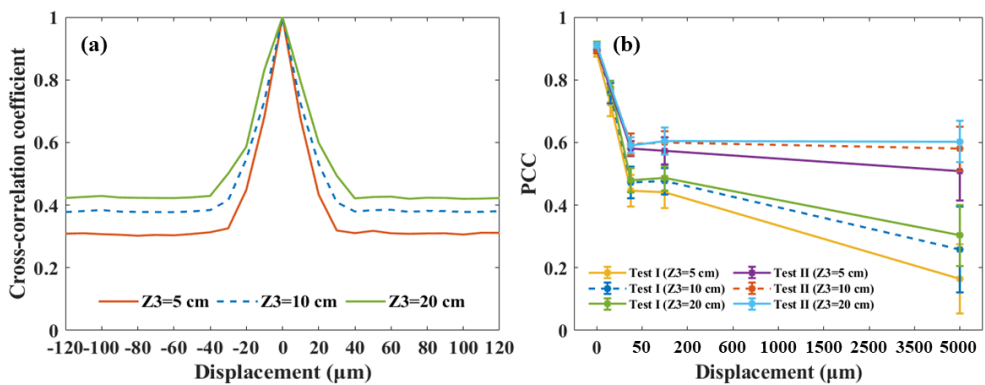

Fig. 6. Comparison of experimental performance with different weights of ballistic photons reaching the detector. (a) Cross-correlation as a function of displacement along the x direction for $z_3 = 5\ cm$, $z_3 = 10\ cm$ and $z_3 = 20\ cm$, respectively. (b) PCC of reconstruction images as a function of displacement along the x direction, with network training data from Position 1 (Test I) and Positions 1-5 (Test II), respectively.

## 4. Discussions and Conclusion

Deep learning shows significant advantages in reconstructing targets through scattering media. Its capability, however, is limited to thin scattering samples, and the generalization of the trained neural networks is case dependent. Aiming to find out why, it is hypothesized in this study that the roles of scattered and ballistic photons in the learning process is different. To confirm that, experiments and simulations were designed and performed by adjusting the weights of ballistic and scattered photons through ground glass diffusers and a random phasemask. It is found that although deep learning can extract images from both scattered and ballistic light, the mechanisms are different: scattering may function as an encryption key and decryption from scattered light is key sensitive, while extraction from ballistic light is stable. Based on this finding, it is shown that the performance of image reconstruction through scattering media can be improved by increasing the weights of ballistic photon in the detection. And furthermore, it is hypothesized and confirmed that the foundation of the generalization capability of trained neural networks for different diffusers can trace back to the contribution of ballistic photons, even though their weights of photon counting in detection are not that significant.

Before conclusion, a few more aspects need to be clarified or discussed. First, the seemingly discrepancy between Refs. [35] and [36], where one can image through a 220-grit diffuser while the other cannot, might be due to the difference of diffusivity for ground glass diffusers from different manufacturers. The one from Edmund is more diffusive than the one from Thorlabs, as shown in the spatial coherence length quantifications, with one being called degree of shift variance along the x or y direction and the other called speckle size characterization, in Figs.2 of both references.

Second, in our study, instead of using different diffusers, different regions of the same diffuser were employed to have better consistency in statistics. That said, there are small variations of the weight of transmitted ballistic photons for different regions. Hence, the network trained with data from 5 positions in Test II yields better generalization capability. Note that the generalization capability comes from ballistic rather than scattered photons. In the captioned scene, scattering functions as an encryption key, which is position and motion sensitive. The trained network inversely mimics the encryptions key, *i.e.*, decrypt the speckle patterns to predict the object. For any particular key, the mimicking accuracy is enhanced if more training data is acquired. For a volumetric turbid medium, as the thickness increases, not only the weight of transmitted ballistic photons reduces, the key also becomes more sensitive to factors such as internal thermal motion of scatterers, medium instability, orientation and size of the incident beam, *etc*. A tiny change may result in a totally different key. Thus, under the same system setup and environment conditions, more encryption keys could appear in the acquisitions of both training and testing data for a thicker medium. In other words, the mimicking accuracy of the trained network for seen keys will deteriorate if training data for each key is reduced, and there is also increased chance that the testing data is encrypted by unseen keys. This is the very reason that through thick scattering media, the performance and generalization capability of the deep learning-based methods reduce due to the lack of sufficient ballistic photons.

Third, it should be clarified that in our setup (Fig. 1), image information is extracted from a diffraction pattern and/or encrypted diffraction pattern. The scenario is a bit different from more general cases where the object plane is imaged onto the CCD plane and image information is extracted from an image and/or encrypted image [35, 36, 48]. The encrypted image is also motion (no matter internal or external motions) sensitive. As thickness increases, the weight of transmitted ballistic photons drops, and the encrypted key of the medium becomes more intricate and delicate. Even through there is no visible changes to the medium, countless keys already encrypt the detected patterns. Then the neural network is not well trained, nor can it reconstruct images even through a trained region. Similar issues occur for methods based on optical phase conjugation, wavefront shaping, and scattering transmission matrix measurement. Basically, even small motions may fail these methods completely.

Last but not the least, the deep learning-based framework can also be applied in holographic 3D particle imaging under different scattering conditions [39]. Nevertheless, most detected backpropagation photons experience only a single scattering event, from where information of a particle is carried. That is to speak, there are significant ballistic photons being detected. The generalization capability of the dynamic synthesis network (DSN) can also trace back to ballistic photons.

In summary, by using deep learning as a probing tool rather than only an image extraction method, we propose and demonstrate that deep learning can extract images from both scattered and ballistic photons, but in different manners. The main difference is that scattering works as an encryption key, and decryption from scattered photons is key sensitive, while extraction from ballistic photons is robust and key insensitive. Moreover, the origin of the generalization capability of trained neural networks for thin scattering layers of identical macroscopic parameters is the better adaptability to fluctuation of ballistic components. These findings add new knowledge to the mechanisms of deep learning for imaging through scattering media and provide guidelines to deepen the penetration optical thickness and to optimize the experimental framework for performance and network generalization enhancement. These may benefit many areas such as biomedical imaging, remote sensing, and underwater detection. Moreover, the study also paves an avenue for deep learning as a probe to explore unknown principles and/or mechanisms in various fields.

**Funding**

The work was supported by National Natural Science Foundation of China (NSFC) (81930048, 81627805), Guangdong Science and Technology Commission (2019A1515011374, 2019BT02X105), Hong Kong Research Grant Council (15217721, R5029-19, C7074-21GF), and Hong Kong Innovation and Technology Commission (GHP/043/19SZ, GHP/044/19GD).

**Acknowledgements**

H. L. conceived the idea and designed the experiment and simulation. X. Z., J. G., Y. G., and C. S. implemented the experiment and simulation. X. Z., P. L., and H. L. analysed and discussed the data. All contribute in writing and revising the manuscript.

**Competing Interests**

The authors declare no conflict of interests.

See Supplement 1 for supporting content.